\documentclass[letterpaper, 10 pt, conference]{ieeeconf}  

\IEEEoverridecommandlockouts                              
\usepackage{graphicx} 
\usepackage{amsmath} 
\usepackage{amssymb}  
\usepackage{subfig}
\usepackage{pgfplotstable} 

\usepackage{booktabs}
\usepackage{footnote}
\makesavenoteenv{table}
\usepackage{hyperref}
\title{\LARGE \bf Deep Convolutional Networks in System Identification}

\author{Carl Andersson$^*$, Ant\^{o}nio H. Ribeiro$^*$, Koen Tiels, Niklas Wahlstr\"om and Thomas B. Sch\"on
\thanks{$^*$Equal contribution.}
\thanks{This work has been supported by the Brazilian agencies CAPES, CNPq and FAPEMIG, by the Swedish Research Council (VR) via the project \emph{NewLEADS -- New Directions in Learning Dynamical Systems} (contract number: 621-2016-06079), and by the Swedish Foundation for Strategic Research (SSF) via the project \emph{ASSEMBLE} (contract number: RIT15-0012). C. Andersson, A. H. Ribeiro, K. Tiels, N. Wahlstr\"om and T. B. Sch\"on are with the Dept. of Information Technology, Uppsala University, 751 05 Uppsala, Sweden. A. Ribeiro is also with the Graduate  Program in  Electrical  Engineering at Universidade Federal de Minas Gerais (UFMG) -
  Av. Ant\^{o}nio Carlos 6627, 31270-901, Belo Horizonte, MG, Brazil
E-mails: \{carl.andersson, antonio.ribeiro, koen.tiels, niklas.wahlstrom, thomas.schon\}@it.uu.se.}}

\begin{document}

\maketitle
\thispagestyle{empty}
\pagestyle{empty}

\begin{abstract}
Recent developments within deep learning are relevant for nonlinear system identification problems. In this paper, we establish connections between the deep learning and the system identification communities. It has recently been shown that convolutional architectures are at least as capable as recurrent architectures when it comes to sequence modeling tasks. Inspired by these results we explore the explicit relationships between the recently proposed temporal convolutional network (TCN) and two classic system identification model structures; Volterra series and block-oriented models. We end the paper with an experimental study where we provide results on two real-world problems, the well-known Silverbox dataset and a newer dataset originating from ground vibration experiments on an F-16 fighter aircraft.
\end{abstract}

\section{Introduction}
\label{sec:introduction}
%
Deep learning has, over the past decade, had a massive impact on several branches of science and engineering, including for example computer vision \cite{KrizhevskySH:2012ImageNet}, speech recognition \cite{Hinton:2012} and natural language processing \cite{MikolovCCD:2013}. While the basic model class---neural networks---has been around for more than 70 years \cite{McCullochP:1943}, there has been quite a few interesting and highly relevant technical developments within the deep learning community that has, to the best of our knowledge, not yet been fully exploited within the system identification community. Just to mention a few of these developments we have; new regularization methods~\cite{iangoodfellow_maxout_2013, srivastava_dropout_2014, liwan_regularization_2013}, new architectures~\cite{simonyan_very_2014, szegedy_going_2015, he_deep_2016}, improved optimization algorithms~\cite{kingma_adam_2014, bottou_optimization_2018, ioffe_batch_2015}, new insights w.r.t.  activation functions~\cite{maas_rectifier_2013, mdzeiler_rectified_2013, he_delving_2015}. Moreover, the capability to significantly increase the depth \cite{simonyan_very_2014,szegedy_going_2015, he_deep_2016} in the models has further  improved the performance. Most of the existing model architectures have been made easily available through high quality open source frameworks~\cite{tensorflow2015-whitepaper, paszke_automatic_2017}, allowing deep learning to be easily implemented, trained, and deployed.

%
The deep learning developments that are most relevant for system identification are probably the ones that can be found under the name of \textit{sequence learning}. Recurrent models, such as recurrent neural networks (RNN) and its extensions which include the long short-term memory (LSTM) \cite{lstm_Hochreiter1997} and the gated recurrent units (GRU) \cite{gru_Gulcehre2014}, have been the standard choice for sequence learning. In the neighbouring area of computer vision, the use of the so-called convolutional neural networks  (CNNs)~\cite{ylecun_gradientbased_1998} has had a very strong impact on tasks such as image classification~\cite{krizhevsky_imagenet_2012}, segmentation~\cite{long_fully_2015} and object detection~\cite{redmon_you_2016a}. Interestingly, it has recently \cite{tcn_Bai2018} been shown that the CNN architecture is highly useful also for sequence learning tasks. More specifically, the so-called temporal CNN (TCN) can match or even outperform the older recurrent architectures in language and music modelling~\cite{tcn_Bai2018, oord_wavenet_2016, dauphin_language_2017}, text-to-speech conversion~\cite{oord_wavenet_2016}, machine translation~\cite{kalchbrenner_neural_2016, gehring_convolutional_2017} and other sequential tasks~\cite{tcn_Bai2018}.  We will, for this reason, focus this paper on making use of TCNs for nonlinear system identification.

%
Neural networks have enjoyed a long and fruitful history \cite{narendra_identification_1990, chen_nonlinear_1990, nonlinear_blackbox_sys_id_overview_1995} also within the system identification community, where they remain a popular choice when it comes to modeling of nonlinear dynamical systems \cite{vargas_improved_2019, dmasti_learning_17, dejesusrubio_stable_2017, kumar_comparative_2019, xqian_generalized_2017}. 

%
We are writing this paper to reinforce the bridge between the system identification and the deep learning communities since we believe that there is a lot to be gained from doing this. We will describe the new TCN model from a system identification point of view (Section~\ref{sec:nns-for-temporal-modeling}). Additionally, we will show that there are indeed interesting connections between the deep TCN structure and 
the Volterra series and the block-oriented model structures commonly used within system identification (Section~\ref{sec:connection-between-comm}). Perhaps most importantly, we will provide experimental results on two real-world problems (the Silverbox \cite{silverbox_wigren2013} and the F-16 \cite{SchoukensN:2017} datasets) and on a toy problem (Section~\ref{sec:numerical-results}).

\section{Neural networks for temporal modeling}
\label{sec:nns-for-temporal-modeling}
The neural network is a universal function approximator~\cite{Hornik1989} with a sequential model structure of the form:
\begin{subequations} \label{eq:NNmap}
\begin{align}
    \hat{y} &= g^{(L)}(z^{(L-1)}), \\
    z^{(l)} &= g^{(l)}(z^{(l-1)}), \quad l=1,\dots,L-1, \\
    z^{(0)} &= x,
\end{align}
\end{subequations}
where $x, z^{(l)}, \hat y$ denotes the input, the hidden variables and the output, respectively. The transformation within each layer is of the form $g^{(l)}(z) = \sigma(W^{(l)} z + b^{(l)})$ consisting of a linear transformation $W^{(l)} z + b^{(l)}$ followed by a scalar nonlinear mapping, $\sigma$, that acts element-wise. 
In the final (output) layer the nonlinearity is usually omitted, i.e. $g^{(L)}(z) = W^{(L)} z + b^{(L)}$.

The neural network parameters $\{W^{(l)}, b^{(l)}\}_{l=1}^{L}$ are usually referred to as the weights $W^{(l)}$ and the bias terms $b^{(l)}$ and they are estimated by minimizing the prediction error ${\frac{1}{N}\sum_{k=1}^N \|\hat y[k]-y[k]\|^2}$ for some training dataset $\{x[k], y[k]\}_{k=1}^N$.

To train deep neural networks with many hidden layers (large $L$) have been proved to be a notoriously hard optimization problem. The challenges includes the risk of getting stuck in bad local minimas, exploding and/or vanishing gradients, and dealing with large-scale datasets. It is only over the past decade that these challenges have been addressed, with improved hardware and algorithms, to the extent that training truly deep neural networks has become feasible. We will very briefly review some of these developments below. Additional information can be found in Appendix A. 

\subsection{Temporal Convolutional Network}
\begin{figure*}
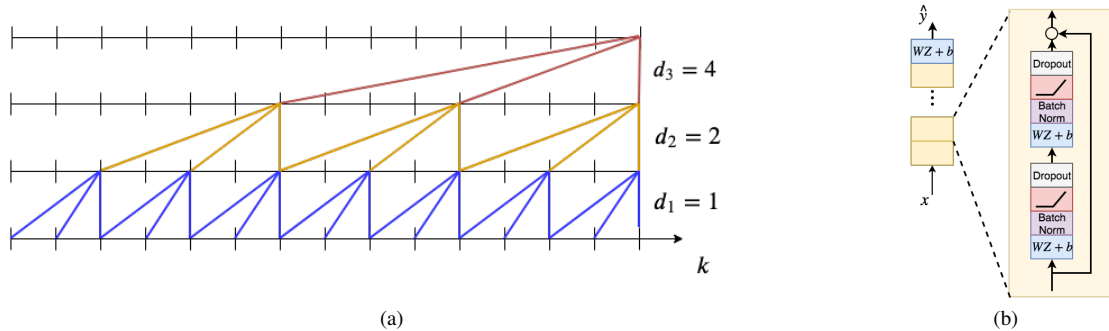

  \centering
    \subfloat[][]{\includegraphics[height=0.17\textheight]{img/dilated_convolutions.png}\label{fig:TCN}}
    ~~~~~~~~~~~~
  \subfloat[][]{\includegraphics[height=0.17\textheight]{img/tcn.png}\label{fig:residual}}
  \caption{Illustration of the temporal convolution network (TCN) with residual blocks. (a) Temporal convolutional network with dilated causal convolutions with dilation factor $d_1=1$, $d_2 = 2$ and  $d_3 = 4$ and kernel size $n=3$. (b) A TCN residual block. Each block consists of two dense layers and an identity (or linear)  map on the skip connection. As illustrated in (a) by connection with the same color, neural network weights are shared within the same layer and invariant to time translations. This reflects the hypothesis we are modeling a time invariant system.}
  \label{fig:simulation_val}
\end{figure*}
As the name suggests, the temporal convolutional network (TCN) is based on convolutions \cite{tcn_Bai2018}. The use of TCNs within a system identification setting can in fact be interpreted as using the nonlinear ARX model as the basic model component:
\begin{align}
\label{eq:NARX}
    \hat y[k+1] = g(x[k],\dots x[k-(n-1)]),
\end{align}
with $x[k] = (u[k],\,\, y[k])$. We will proceed with this interpretation, where~\eqref{eq:NARX} would correspond to a one-layer TCN model.

A full TCN can be understood as a sequential construction of several nonlinear ARX models stacked on top of each other:
\begin{subequations} \label{eq:TCN}
\begin{align}
    \hat y[k+1] & = g^{(L)}(Z^{(L-1)}[k]), \\
    z^{(l)}[k] &= g^{(l)}(Z^{(l-1)}[k]), \quad l=1,\dots,L-1, \\
    z^{(0)}[k] &=  x[k],
\end{align}
where:
\begin{align*}
Z^{(l\hspace{-0.5mm}-\hspace{-0.5mm}1)}[k] = \Big(z^{(l\hspace{-0.5mm}-\hspace{-0.5mm}1)}[k], z^{(l\hspace{-0.5mm}-\hspace{-0.5mm}1)}[k\hspace{-1mm}-\hspace{-1mm}d_l],\dots, z^{(l\hspace{-0.5mm}-\hspace{-0.5mm}1)}[k\hspace{-1mm}-\hspace{-1mm} (n\hspace{-1mm}-\hspace{-1mm}1) d_l]\Big).
\end{align*}
\end{subequations}
The number of layers $L$, the size of each intermediate layer $z^{(l)}[k]$, and the model order~$n$, are all design choices determined by the user. This will also determine the number of parameters included in the model.

For each layer, we optionally introduce a \emph{dilation factor}~$d_l$. With $d_l = 1$ we recover the standard nonlinear ARX model in each layer. With $d_l>1$ the corresponding output of that layer can represent a wider range of past time instances. The effective memory of that layer will be $(n-1)d_l$. Typically the dilation factor is chosen to increase exponentially with the number of layers, for example $d_l=2^{(l-1)}$, see Fig.~\ref{fig:TCN}. If we assume that we have the same number of parameters in each layer, the memory will increase exponentially, not only with the number of layers, but also with the number of parameters in the model. This is a very attractive but yet uncommon property for system identification models present in the literature. 

Each layer in a TCN can also be seen as dilated causal convolution where $n$ would be the kernel size and $\text{dim}(z^{(l)}[k])$ the number of channels in layer $l$. These convolutions can be efficiently implemented in a vectorized manner where many computations are reused across the different time steps $k$. Analogously to what is done in convolutional neural networks we use zero-padding for $z^{(l)}[k]$ where $k < 1$. We refer to \cite{tcn_Bai2018} for a presentation of TCN based on convolutions.

\subsection{Residual Blocks}
A residual block is a combination of possibly several layers together with a skip connection
\begin{align}
z^{(l+p)} = \mathcal{F}(z^{(l)}) + z^{(l)},
\end{align}
where the skip connection adds the value from the input of the block to its output. The purpose of the residual block is to let the layers learn deviations from the identity map rather than the whole transformation. This property is beneficial, especially in deep networks where we want the information to gradually change from the input to the output as we proceed through the layers. There is also some evidence that this makes it easier to train deeper neural networks~\cite{he_deep_2016}.

We employ residual blocks in our models by following the model structure in \cite{tcn_Bai2018}. Each block consist of one skip connection and two linear mappings, each of them followed by batch normalization~\cite{ioffe_batch_2015}, activation function and dropout regularization~\cite{srivastava_dropout_2014}. See Fig.~\ref{fig:simulation_val}b for a visual description and Appendix A for a brief explanation of batch normalization and regularization methods.  For both mappings a common dilation factor is used and hence the whole block can be seen as one of the layers $g^{(l)}(z)$ in the TCN model~\eqref{eq:TCN}. Note that the skip connection only passes $z^{(l-1)}[k]$ to the next layer and not the whole $Z^{(l-1)}[k]$ for each time instance~$k$. In cases where $z^{(l-1)}[k]$ and $z^{(l)}[k]$ are of different dimensions, a linear mapping is used between them. The coefficients of this linear mapping are also learned during training.

\section{Connections to system identification}
\label{sec:connection-between-comm}
This section describes equivalences between the basic TCN architecture (i.e.~without dilated convolutions and skip connections) and models in the system identification community, namely Volterra series and block-oriented models. The discussion is limited to the nonlinear FIR case (where $x=u$) instead of the more general NARX case ($x=(u,y)$) considered in \eqref{eq:NARX}, and to single input single output systems.

\subsection{Connection with Volterra series}

%
%
A Volterra series \cite{Schetzen2006} can be considered as a Taylor series with memory. It is essentially a polynomial in (delayed) inputs $u[k], u[k-1], \ldots$. Alternatively, a Volterra series can be considered as a nonlinear generalization of the impulse response $h_1[\tau]$. The output of a Volterra series is obtained using higher-order convolutions of the input with the Volterra kernels $h_d[\tau_1, \ldots, \tau_d]$ for $d=0, 1, \ldots, D$. These kernels are the polynomial coefficients in the Taylor series.

%
%
The basic TCN architecture is essentially the same as the time delay neural network (TDNN) in \cite{Waibel1989}, except for the zero padding \cite{tcn_Bai2018} and the use of ReLU activations instead of sigmoids. The TDNN has been shown to be equivalent to an infinite-degree ($D \rightarrow \infty$) Volterra series in \cite{WrayGreen1994}. This connection is made explicit in \cite{WrayGreen1994} by showing how to compute the Volterra kernels from the estimated network weights $W$. The key ingredient is to use a Taylor series expansion of the activation functions $\sigma$ either around zero (the bias term~$b$ is then considered part of the activation function) or alternatively around the bias values if the Taylor series around zero does not converge for example.

\subsection{Connection with block-oriented models}

%
%
Block-oriented models \cite{GiriBai2010,SchoukensM2017} combine linear time-invariant (LTI) subsystems (or blocks) and nonlinear static (i.e.~memoryless) blocks. For example, a Wiener model consists of the cascade of an LTI block and a nonlinear static block. For a Hammerstein model, the order is reversed: it is a nonlinear block followed by an LTI block. Generalizations of these simple structures are obtained by putting more blocks in series (as in \cite{WillsNinness2012} for Hammerstein systems) or in parallel branches (as in \cite{schoukens_parametric_2015} for Wiener-Hammerstein systems) and/or to consider multivariate nonlinear blocks.

%
%
%
%
A multi-layer basic TCN can be considered as cascading parallel Wiener models, one for each hidden layer, that have multivariate nonlinear blocks consisting of the activation functions (including the bias). The linear output layer corresponds to adding FIR filters at the end of each parallel branch.
The layers can be squeezed together to form less but larger layers (cf.~squeezing together the sandwich model discussed in \cite{Palm1979}). This is so since the dynamics consist of time delays and time delays can be placed before or after a static nonlinear function without changing the resulting output ($q^{-1}\sigma(z[k]) = \sigma(q^{-1}z[k]) = \sigma(z[k-1])$). The TCN model could be squeezed down to a parallel Wiener model.

\subsection{Conclusion}

%
%
The basic TCN architecture is equivalent to Volterra series and parallel Wiener models. They are thus all universal approximators for time-invariant systems with fading memory \cite{BoydChua1985}. This equivalence does \textit{not} mean that all these model structures can be trained with equal ease and will perform equally well in all identification tasks. For example, a Volterra series uses polynomial basis functions, whereas TDNNs use sigmoids and TCNs use ReLU activation functions. Depending on the system at hand, one basis function might be better suited than another to avoid bad local minima and/or to obtain both an accurate and sparse representation.





\section{Numerical results}
\label{sec:numerical-results}

We now present the performance of the TCN model on three system identification problems. We compare this model with the classical NARX Multilayer Perceptron (MLP) network with two layers and with the Long Short-Term Memory (LSTM) network. When available, results from other papers on the same problem are also presented.

We make a distinction between \textit{training}, \textit{validation} and \textit{test} datasets. The \textit{training} dataset is used for estimating the parameters. The performance in the \textit{validation} data is used as the  early stopping criteria for the optimization algorithm and for choosing the best hyper-parameters (i.e. neural network number of layers, number of hidden nodes, optimization parameter, and so on). The \textit{test} data allows us to assess the model performance on unseen data.  Since the major goal of the first example is to compare different hyper-parameter choices we do not use a \textit{test} set.

In all the cases, the neural network parameters are estimated by minimizing the mean square error using the Adam optimizer~\cite{kingma_adam_2014} with default parameters and an initial learning rate of $\text{lr} = 0.001$. The learning rate is reduced whenever the validation loss does not improve for 10 consecutive epochs. 

We use the Root Mean Square Error (${\text{RMSE} =  \sqrt{\frac{1}{N}\sum_{k=1}^N\|\hat{y}[k] - y[k]\|^2}}$) as metric for comparing
the different methods in the validation and test data. Throughout the text we will make clear when the predicted output $\hat{y}$ is computed through the free-run simulation of the model and when it is computed through the one-step-ahead prediction.

The code for reproducing the examples is available at \href{https://github.com/antonior92/sysid-neuralnet}{https://github.com/antonior92/sysid-neuralnet}. Additional information about the hyperparameters and training time can be found in Appendix B.

\subsection{Example 1: Nonlinear toy problem}
\label{sec:example_1}
\begin{figure}[h]
  \centering
  \includegraphics[width=0.5\textwidth]{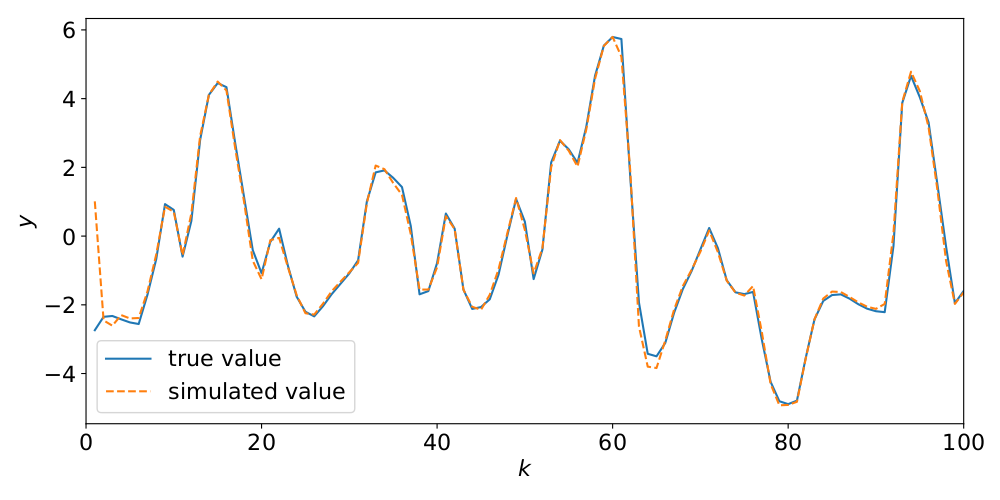}
      \caption{\textbf{(Example 1)} Displays $100$ samples of the free-run simulation TCN model vs the simulation of the true system. The kernel size for the causal convolutions is $2$, the dropout rate is $0$, it has $5$ convolutional layers and a dilation rate of~$1$. The training set has $20$ batches of $100$ samples and was generated with~(\ref{eq:ex_nonlinear_system}) for $v$ and $w$ white Gaussian noise with standard deviations $\sigma_v = 0.3$ and $\sigma_w = 0.3$. The validation set has $2$ batches of $100$ samples. For both, the input $u$ is randomly generated with a standard Gaussian distribution, each randomly generated value held for $5$ samples.}
  \label{fig:simulation_val_chen}
\end{figure}

The nonlinear system~\cite{chen_nonlinear_1990}:
\begin{eqnarray}
  y^*[k] &=& (0.8-0.5e^{-y^*[k-1]^2})y^*[k-1]- \nonumber\\
    &&(0.3+0.9e^{-y^*[k-1]^2})y^*[k-2]+u[k-1]+ \nonumber\\
    &&0.2u[k-2]+0.1u[k-1]u[k-2] + v[k], \nonumber\\
y[k] &=& y^*[k] + w[k], \label{eq:ex_nonlinear_system}
\end{eqnarray}
was simulated and the generated dataset was used to build neural network models. Fig.~\ref{fig:simulation_val_chen} shows the validation results for a model obtained for a training and validation set generated with white Gaussian process noise~$v$ and measurement noise~$w$. In this section, we repeat this same experiment for different neural network architectures, with different noise levels and different training set sizes~$N$. 

The best results for each neural network architecture on the validation set are compared in Table~\ref{tab:performance_chen}. It is interesting to see that when few samples (${N=500}$) are available for training, the TCN performs the best among the different architectures. On the other hand, when there is more data (${N=8\thinspace000}$) the other architectures gives the best performance.

\begin{table*}
    \caption{\textbf{(Example 1)} One-step-ahead RMSE on the validation set for the models (MLP, LSTM and TCN) trained on datasets generated with: different noise levels ($\sigma$) and  lengths ($N$). The standard deviation of both the process noise~$v$ and the measurement noise~$w$ is denoted by~$\sigma$. We report only the best results among all hyper-parameters and architecture choices we have tried out for each entry.}
    \label{tab:performance_chen}
    \centering
    \pgfkeys{/pgf/number format/.cd,fixed,precision=3, fixed zerofill=true}
    \makebox[\textwidth][c]{ 
    \begin{filecontents*}{tables/optimal_models_chen.csv}
,i,i,i,i,i,i,i,i,i
n_batches,5,5,5,20,20,20,80,80,80
model,lstm,mlp,tcn,lstm,mlp,tcn,lstm,mlp,tcn
noise_levels,,,,,,,,,
0.000,0.362,0.270,0.254,0.245,0.204,0.196,0.165,0.154,0.159
0.300,0.712,0.645,0.607,0.602,0.586,0.558,0.549,0.561,0.551
0.600,1.183,1.160,1.094,1.105,1.070,1.066,1.038,1.052,1.043
\end{filecontents*}
\pgfplotstabletypeset[
    col sep=comma,
    skip first n=3,
    display columns/0/.style={column name = $\sigma$, /pgf/number format/.cd,fixed,precision=1},
    display columns/1/.style={column name = LSTM, column type=|c},
    display columns/2/.style={column name = MLP},
    display columns/3/.style={column name = TCN},
    display columns/4/.style={column name = LSTM, column type=|c},
    display columns/5/.style={column name = MLP},
    display columns/6/.style={column name = TCN},
    display columns/7/.style={column name = LSTM, column type=|c},
    display columns/8/.style={column name = MLP},
    display columns/9/.style={column name = TCN},
    every head row/.style={
    before row={
        \toprule
        \multicolumn{1}{c}{}& \multicolumn{3}{c}{N=500} & \multicolumn{3}{c}{N=2\thinspace000} & \multicolumn{3}{c}{N=8\thinspace000}\\
        },
      after row=\midrule,
      },
      every last row/.style={
      after row=\bottomrule},
      every row 0 column 3/.style={postproc cell content/.style={@cell content=\textbf{##1}}},
      every row 0 column 6/.style={postproc cell content/.style={@cell content=\textbf{##1}}},
      every row 0 column 8/.style={postproc cell content/.style={@cell content=\textbf{##1}}},
      every row 1 column 3/.style={postproc cell content/.style={@cell content=\textbf{##1}}},
      every row 1 column 6/.style={postproc cell content/.style={@cell content=\textbf{##1}}},
      every row 1 column 7/.style={postproc cell content/.style={@cell content=\textbf{##1}}},
      every row 2 column 3/.style={postproc cell content/.style={@cell content=\textbf{##1}}},
      every row 2 column 6/.style={postproc cell content/.style={@cell content=\textbf{##1}}},
      every row 2 column 7/.style={postproc cell content/.style={@cell content=\textbf{##1}}},
    ]{tables/optimal_models_chen.csv}
    }
\end{table*}

\begin{figure*}
    \centering
    \subfloat[][]{\includegraphics[width=0.25\textwidth]{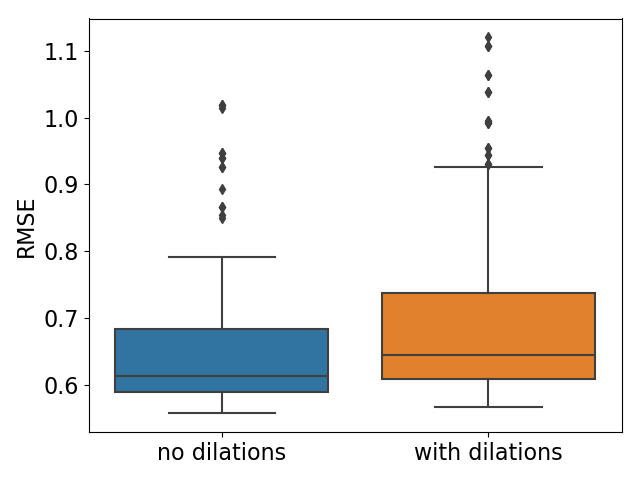}}
    \subfloat[][]{\includegraphics[width=0.25\textwidth]{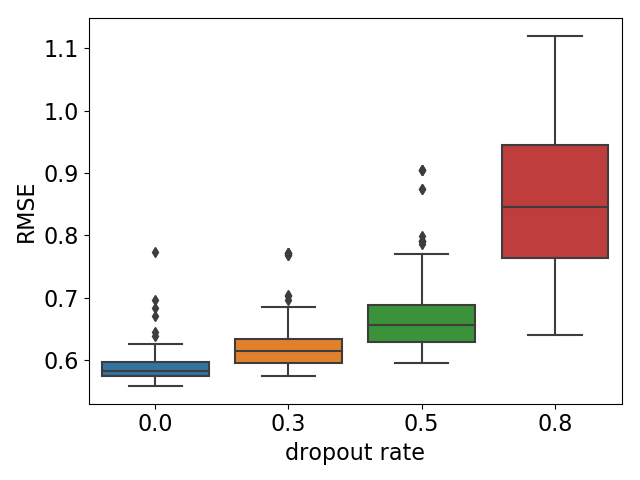}}
    \subfloat[][]{\includegraphics[width=0.25\textwidth]{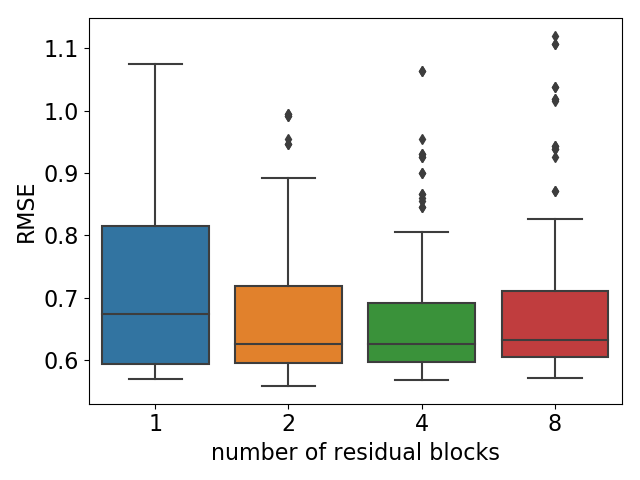}}
    \subfloat[][]{\includegraphics[width=0.25\textwidth]{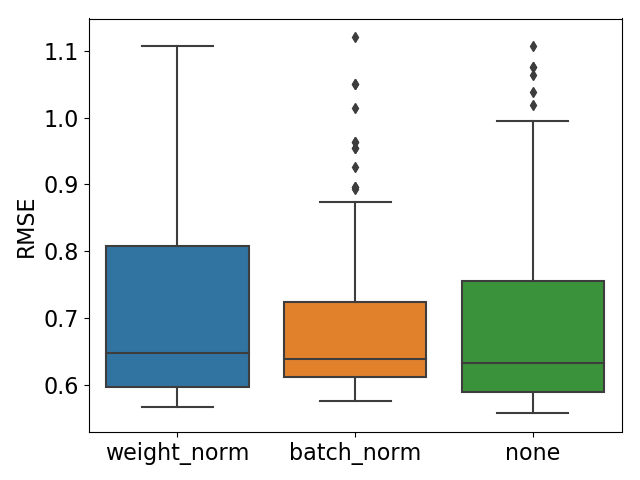}}
    \caption{\textbf{(Example 1)} Box plots showing how different design choices affect the performance of the TCN for noise standard deviation $\sigma=0.3$ and training data length $N=2\thinspace000$. On the $y$-axis the one-step-ahead RMSE on the validation set is displayed, and on the $x$-axis we have: in (a) the presence or absence of dilations; in (b) the dropout rate \{0.0, 0.3, 0.5, 0.8\}; in (c) the number of residual blocks \{1, 2, 4, 8\}; and, in (d) if \textit{batch norm}, \textit{weight norm} or nothing is used for normalizing the output of each convolutional layer. The variation in performance for the box plot quartiles is achieved through the variation for all the other hyper-parameters not fixed by the hyper-parameter choice indicated on the $x$-axis.}
    \label{fig:performance_tcn_hyperparameters}
\end{figure*}

Fig.~\ref{fig:performance_tcn_hyperparameters} shows how different hyper-parameter choices impact the performance of the TCN. We note that standard deep learning techniques such as dropout, batch normalization and weight normalization did not improve performance. The use of dropout actually hurts the model performance on the validation set. Furthermore, increasing the depth of the neural network does not actually improve its performance and the TCN yields better results in the training set without the use of dilations, which makes sense considering that this model does not require a long memory since the data were generated by a system of order 2.

\subsection{Example 2: Silverbox}
The Silverbox is an electronic circuit that mimics the input/output behavior of a mass-spring-damper with a cubic hardening spring. A benchmark dataset is available through \cite{silverbox_wigren2013}.\footnote{Data available for download at:\\ \href{http://www.nonlinearbenchmark.org/\#Silverbox}{http://www.nonlinearbenchmark.org/\#Silverbox}}

The training and validation input consists of 10 realizations of a random-phase multisine.  Since the process noise and measurement noise is almost nonexistent in this system, we use all the multisine realizations for training data, simply training until convergence. 

\begin{figure}[h]
  \centering
  \includegraphics[width=0.5\textwidth]{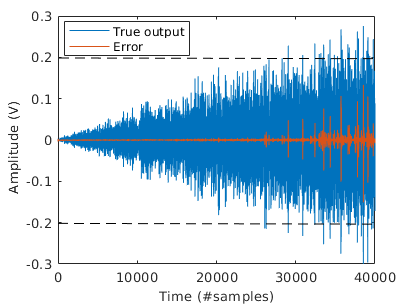}
  \caption{\textbf{(Example 2)} The true output and the prediction error of the TCN model in free-run simulation for the Silverbox data. The model needs to extrapolate  approximately outside the region $\pm 0.2$ marked by the dashed lines.}
  \label{fig:silverbox_extrapolation}
\end{figure}

The test input consists of 40\thinspace400 samples of a Gaussian noise with a linearly increasing amplitude. This leads to the variance of the test output being larger than the variance seen in the training and validation dataset in the last third of the test data, hence the model needs to extrapolate in this region. Fig.~\ref{fig:silverbox_extrapolation} visualizes this extrapolation problem and Table~\ref{tab:silverbox_partial} shows the RMSE only in the region where no extrapolation is needed. The corresponding RMSE for the full dataset is presented in Table~\ref{tab:silverbox_full}. Similarly to Section~\ref{sec:example_1} we found that the TCN did not benefit from the standard deep learning techniques such as dropout and batch normalization. We also see that the LSTM outperforms the MLP and the TCN suggesting the Silverbox data is large enough to benefit of the increased complexity of the LSTM. 

\begin{table}
\centering
\caption{\textbf{(Example 2)} Free-run simulation results for the Silverbox example on part of the test data (avoiding extrapolation).}
\label{tab:silverbox_partial}
{\renewcommand{\arraystretch}{0.5}
\begin{tabular}{@{}llll@{}}
\toprule
RMSE (mV) & Which samples & Approach & Reference \\
\midrule
0.7 & first 25\thinspace000 & Local Linear State Space & \cite{Verdult2004} \\
0.24 & first 30\thinspace000 & NLSS with sigmoids & \cite{Marconato2012} \\
1.9 & 400 to 30\thinspace000 & Wiener-Schetzen & \cite{Tiels2015} \\
\midrule
0.31 & first 25\thinspace000 & LSTM & this paper \\
0.58 & first 30\thinspace000 & LSTM & this paper \\
0.75 & first 25\thinspace000 & MLP & this paper \\
0.95 & first 30\thinspace000 & MLP & this paper \\
0.75 & first 25\thinspace000 & TCN & this paper \\
1.16 & first 30\thinspace000 & TCN & this paper \\
\bottomrule
\end{tabular}
}
\end{table}

\begin{table}
\centering
\caption{\textbf{(Example 2)} Free-run simulation results for the Silverbox example on the full test data. \footnotesize{($^*$Computed from FIT=92.2886\%}).}
\label{tab:silverbox_full}
{\renewcommand{\arraystretch}{0.5}
\begin{tabular}{@{}lll@{}}
\toprule
RMSE (mV) & Approach & Reference \\
\midrule
0.96 & Physical block-oriented & \cite{Hjalmarsson2004} \\
0.38 & Physical block-oriented & \cite{Paduart2004} \\
0.30 & Nonlinear ARX & \cite{Ljung2004} \\
0.32 & LSSVM with NARX & \cite{Espinoza2004} \\
1.3 & Local Linear State Space & \cite{Verdult2004} \\
0.26 & PNLSS & \cite{Paduart2008} \\
13.7 & Best Linear Approximation & \cite{Paduart2008} \\
0.35 & Poly-LFR & \cite{VanMulders2013} \\
0.34 & NLSS with sigmoids & \cite{Marconato2012} \\
0.27 & PWL-LSSVM with PWL-NARX & \cite{Espinoza2005} \\
7.8 & MLP-ANN & \cite{Sragner2004} \\
4.08$^*$ & Piece-wise affine LFR & \cite{Pepona2011} \\
9.1 & Extended fuzzy logic & \cite{Sabahi2016} \\
9.2 & Wiener-Schetzen & \cite{Tiels2015} \\
\midrule
3.98 & LSTM & this paper \\
4.08 & MLP & this paper \\
4.88  & TCN & this paper \\
\bottomrule
\end{tabular}
}
\end{table}

\subsection{Example 3: F-16 ground vibration test}

\begin{figure}
    \centering
    \includegraphics[width=0.25\textwidth]{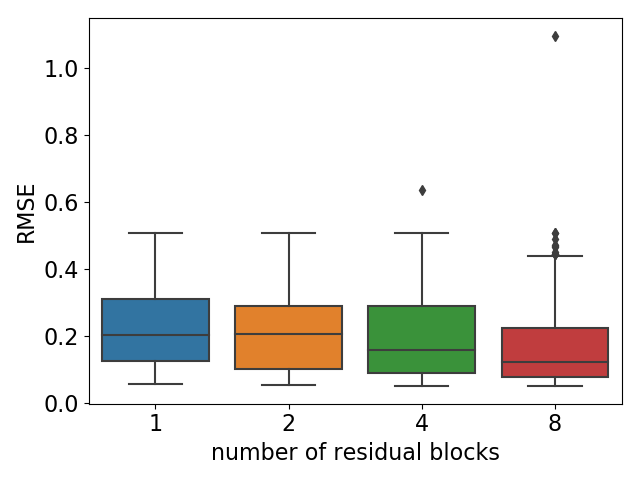}
    \caption{\textbf{(Example 3)} Box plot showing how different depths of the neural network affects the performance of the TCN. Should be interpreted in the same way as Fig.~\ref{fig:performance_tcn_hyperparameters}. }
    \label{fig:performance_tcn_hyperparameters_f16}
\end{figure}

The F-16 vibration test was conducted on a real F-16 fighter equipped with dummy ordnances and accelerometers to measure the structural dynamics of the interface between the aircraft and the ordnance. A shaker mounted on the wing was used to generate multisine inputs to measure this dynamics. We used the multisine realizations with random frequency grid with 49.0 N RMS amplitude~\cite{SchoukensN:2017} for training, validating  and testing the model.\footnote{Data available for download at: \\ \href{http://www.nonlinearbenchmark.org/\#F16}{http://www.nonlinearbenchmark.org/\#F16}}

We trained the TCN, MLP and LSTM networks for all the same configurations used in Example~1. The analysis of the different architecture choices for the TCN in the validation set again reveals that common deep learning techniques such as dropout, batch normalization, weight normalization or the use of dilations do not improve performance. The major difference here is that the use of a deeper neural network actually outperforms shallow neural networks (Fig.~\ref{fig:performance_tcn_hyperparameters_f16}).

The best results for each neural network architecture are compared in Table~\ref{tab:f16_avg} for free-run simulation and one-step-ahead prediction. The results are averaged over the 3 outputs. The TCN performs similar to the LSTM and the MLP.

\begin{figure}
    \centering
    \includegraphics[width=0.38\textwidth]{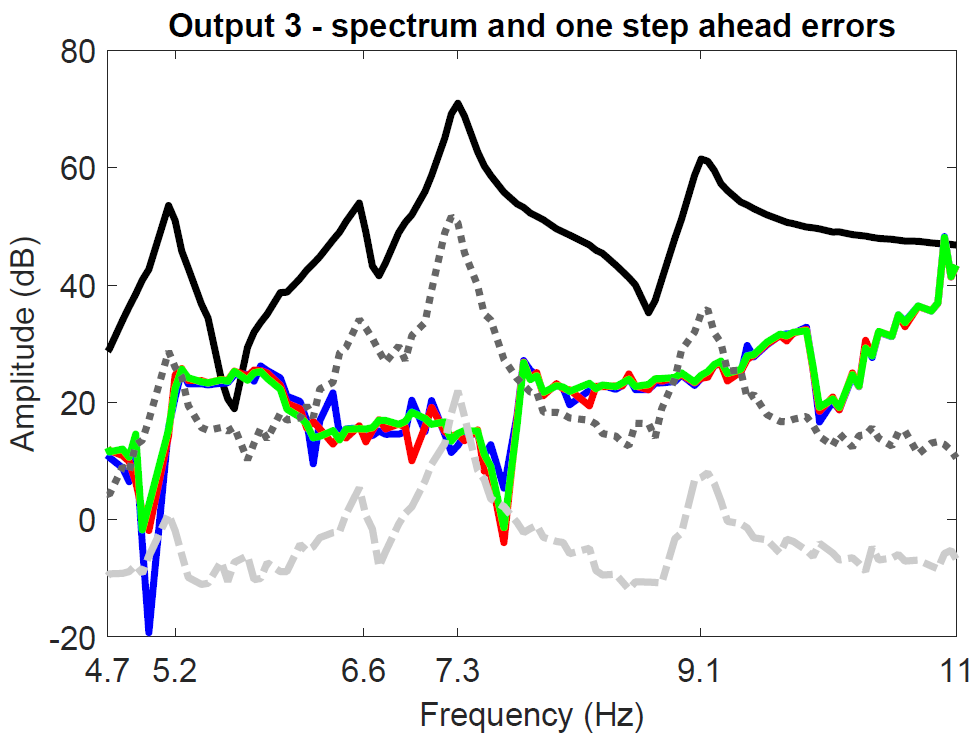}
    \caption{\textbf{(Example 3)} In one-step-ahead prediction mode, all tested model structures perform similar. The error is close to the noise floor around the main resonance at 7.3 Hz. (plot only at excited frequencies in $[4.7,11]$ Hz; true output spectrum in black, noise distortion in grey dash-dotted line, total distortion (= noise + nonlinear distortions) in grey dotted line, error LSTM in green, error MLP in blue, and error TCN in red)}
    \label{fig:f16_osa_output3_zoom}
\end{figure}

An earlier attempt on this dataset with a polynomial nonlinear state-space (PNLSS) model is reported in \cite{Tiels2017}. Due to the large amount of data and the large model order, the complexity of the PNLSS model had to be reduced and the optimization had to be focused in a limited frequency band (4.7 to 11 Hz). That PNLSS model only slightly improved on a linear model. Compared to that, the LSTM, MLP, and TCN perform better, also in the frequency band 4.7 to 11 Hz. This can be observed in Fig.~\ref{fig:f16_osa_output3_zoom}, which compare the errors of these models with the noise floor and the total distortion level (= noise + nonlinear distortions), computed using the robust method \cite{Pintelon2012}. Around the main resonance at 7.3 Hz (the wing-torsion mode \cite{SchoukensN:2017}), the errors of the neural networks are significantly smaller than the total distortion level, indicating that the models do capture significant nonlinear behavior.
Similar results are obtained in free-run simulation (not shown here).
In contrast to the PNLSS models, the neural networks did not have to be reduced in complexity. Due to the mini-batch gradient descent, it is possible to train complex models on large amounts of data. 

\begin{table}[]
    \centering
    \caption{\textbf{(Example 3)} RMSE for free-run simulation and one-step-ahead prediction for the F16 example averaged over the 3 outputs. The average RMS value of the 3 outputs is 1.0046.}
    \label{tab:f16_avg}
    {\renewcommand{\arraystretch}{0.5}
    \begin{tabular}{@{}llll@{}}
        \toprule
        Mode & LSTM & MLP & TCN \\
        \midrule
        Free-run simulation & 0.74 & 0.48 & 0.63 \\
        One-step-ahead prediction & 0.023 & 0.045 & 0.034 \\
        \bottomrule
    \end{tabular}
    }
\end{table}



\section{Conclusion and Future Work}
\label{sec:conclusion}


In this paper we applied recent deep learning methods to standard system identification benchmarks. Our initial results indicate that these models have potential to provide good results in system identification problems, even if this requires us to rethink how to train and regularize these models. Indeed, methods which are used in traditional deep learning settings do not always improve the performance. For example, dropout did not yield better results in any of the problems. Neither did the long memory offered by the dilation factor in TCNs offer any improvement, which is most likely due to the fact that these problems have a relatively short and exponentially decaying memory, as most dynamical systems do. Other findings are that TCNs work well also for small datasets and that LSTMs did show a good overall performance despite being very rarely applied to system identification problems.

Causal convolutions are effectively similar to NARX models and share statistical properties with this class of models. Hence, they are also expected to be biased for settings where the noise is not white. This could justify the limitations of TCNs observed in our experiments. Extending TCNs to handle situations where the data is contaminated with non-white noise seems to be a promising direction in improving the performance of these models. Furthermore, both LSTMs and the dilated TCNs are designed to work well for data with long memory dependencies. Therefore it would be interesting to apply these models to system identification problems where such long term memory is actually needed, e.g. switched systems, or to study if the long-term memory can be translated into accurate long-term predictions, which could have interesting applications in a model predictive control setting.



\bibliographystyle{IEEEtran}
\bibliography{refs,antonio_refs} 


\newpage
\appendix

\subsection{Neural networks common practices}
\subsubsection{Regularization}
Similar to other approaches within system identification, L2- and L1-regularization are commonly used to reduce the flexibility of a model and hence avoid overfitting. A number of other regularization techniques have also appeared more specialized to neural networks. One of them is the dropout~\cite{srivastava_dropout_2014} which is a technique where a random subset of the hidden units in each layer is set to zero during training. New random subsets are drawn and set to zero in each optimization step which effectively means that a random subnetwork is trained during each iteration.

Data augmentation is very common in classification problems and it can also be interpreted as a regularization technique. It is used to artificially increase the training dataset by utilizing the fact that the class is invariant under some transformation of the input (e.g. translation for image) or in the presence of some low intensity noise (e.g. salt pepper noise for images).

Finally, early stopping is a pragmatic approach in which, as the name suggests, the optimization algorithm is interrupted before convergence. The stopping point is chosen as the point where a validation error is minimized. Hence, it avoids overfitting and can as such also be interpreted as a regularization technique.

\subsubsection{Batch Normalization}
Before training a neural network, the inputs are commonly normalized by subtracting the mean and dividing by the variance. The purpose of this is to avoid early saturation of the activation function and assuring that values in the proceeding layers are within the same dynamic range. In deep networks it is beneficial to not only normalize the input layer, but also the intermediate hidden layers. This idea is exploited in batch normalization \cite{ioffe_batch_2015} which, in addition to this normalization, introduces scaling parameters~$\gamma$ and a shift~$\beta$  to be learned during training. The output of the layer is then:
\begin{align}
    \tilde z^{(l)}[k] = \gamma \bar z^{(l)}[k] + \beta.
\end{align}
where  $\bar z^{(l)}[k]$ is normalized version of layer $l$ output. The parameters $\gamma$ and $\beta$ will be trained jointly with all other parameters of the network. Batch normalization has become very popular in deep learning models.

An alternative to batch normalization is weight normalization which is, essentially, a reparametrization of the weight matrix, decoupling the magnitude and direction of the weights~\cite{weight_norm_Salimans2016}.

\subsubsection{Optimization Algorithms}
\label{sec:optimization-algorithms}
Neural networks are trained using gradient-based optimization methods. At each iteration only a random subset of the training data is used to compute the gradient and update the parameters. This is called mini-batch gradient descent and is a crucial component for efficient training of a neural network when the dataset is large.

Multiple extensions to mini-batch gradient descent have been proposed to make the learning more efficient. Momentum \cite{qian1999momentum} applies a first order low-pass filter to the stochastic gradients to compensate for the noise introduced by the random sub-sampling. RMSprop \cite{Tieleman2012} uses a low-passed version of the squared gradients to scale the learning rate in the different dimensions. One of the most popular optimization method today is referred to as Adam \cite{kingma_adam_2014} which basically amounts to using RMSprop with momentum.

\subsection{Hyperparameter search and training time}

All examples run with hardware acceleration provided by a single graphical processing unit (GPU). We run different experiments in machines with different configurations so the times are not directly comparable. Some of these machines have a NVIDIA Titan V and others a NVIDIA GTX 1080TI.

A in depth analysis of the training time is beyond the scope of this paper. The idea is to provide some basic notion of how much time is needed to run the neural network and the computational cost of doing hyperparameter search. For example 2, we provide the total time needed to do the hyperparameter search. It should be noticed, however, that grid search is an inefficient  procedure. We choose to use it in order to study the effect of hyperparameters, rather than because of its efficiency. And, for example 3, we provide the total time for training the neural network with the best possible configuration.

\subsubsection{Nonlinear toy problem}
We used grid search for finding the hyperparameters. In each possible training configuration, we have trained the TCN for all possible combinations of: number of hidden layers in \{16, 32, 64, 128, 256\}; dropout rate in \{0.0, 0.3, 0.5, 0.8\}; number of residual blocks in \{1, 2, 4, 8\}; for the kernel size  in \{2, 4, 8, 16\}; for the presence or absence of dilations; and, for the use of \textit{batch norm}, \textit{weight norm} or nothing after each convolutional layer. We have trained the MLP for all combination of: number of hidden layers in \{16, 32, 64, 128, 256\}; model order $n$ in \{2, 4, 8, 16, 32, 64, 128\}; activation function in \{ReLU, sigmoid\}. Finally, we trained the LSTM for all combinations of: number of hidden layers in \{16, 32, 64, 128\}; number of stacked LSTM layers in \{1, 2, 3\}; dropout rate in \{0.0, 0.3, 0.5, 0.8\}. The best hyperparameters for each configuration are described in Table~\ref{tab:hyperparameters_chen}. For the TCN, it is better (in all configurations) to use no dilation, no normalization and kernel size equals to 2, hence these hyperparameters are omitted from the table.

\begin{table*}
    \caption{\textbf{(Example 1)} Best model hyperparameters for: different noise levels ($\sigma$) and  lengths ($N$). The standard deviation of both the process noise~$v$ and the measurement noise~$w$ is denoted by~$\sigma$.}
    \label{tab:hyperparameters_chen}
    \centering
    \pgfkeys{/pgf/number format/.cd,fixed,precision=0, fixed zerofill=true}
    \subfloat[][\textbf{TCN:}  The hyperparameters are the dropout rate (drop.),  the number of layers (n. layers  and the number of hidden layers (h. size).]{
    \begin{filecontents*}{tables/hyperparam_tcn_chen.csv}
,i,i,i,i,i,i,i,i,i
n_batches,5,5,5,20,20,20,80,80,80
hyperparam,dropout,hidden size,n layers,dropout,hidden size,n layers,dropout,hidden size,n layers
noise_levels,,,,,,,,,
0.000,0.0,2,256,0.0,2,64,0.0,4,32
0.300,0.3,8,128,0.0,2,128,0.0,8,16
0.600,0.0,8,256,0.0,1,64,0.0,8,16
\end{filecontents*}
\pgfplotstabletypeset[
    col sep=comma,
    skip first n=3,
    display columns/0/.style={column name = $\sigma$, /pgf/number format/.cd,fixed,precision=1},
    display columns/1/.style={column name = drop., precision=1, column type=|c},
    display columns/2/.style={column name =  n. layers},
    display columns/3/.style={column name = h. size},
    display columns/4/.style={column name = drop., precision=1, column type=|c},
    display columns/5/.style={column name =  n. layers},
    display columns/6/.style={column name = h. size},
    display columns/7/.style={column name = drop., precision=1, column type=|c},
    display columns/8/.style={column name =  n. layers},
    display columns/9/.style={column name = h. size},
    every head row/.style={
    before row={
        \toprule
        \multicolumn{1}{c}{}& \multicolumn{3}{c}{N=500} & \multicolumn{3}{c}{N=2\thinspace000} & \multicolumn{3}{c}{N=8\thinspace000}\\
        },
      after row=\midrule,
      },
      every last row/.style={
      after row=\bottomrule},
    ]{tables/hyperparam_tcn_chen.csv}
    }\\
    \subfloat[][\textbf{MLP:} The hyperparameters are the activation function (activ. fun.), the number of hidden layers (h. size) and the model order $n$.]{
    \begin{filecontents*}{tables/hyperparam_mlp_chen.csv}
,i,i,i,i,i,i,i,i,i
n_batches,5,5,5,20,20,20,80,80,80
hyperparam,activation fun,hidden size,model order,activation fun,hidden size,model order,activation fun,hidden size,model order
noise_levels,,,,,,,,,
0.000,relu,128,2,relu,128,3,relu,256,3
0.300,relu,256,2,sigmoid,64,3,relu,128,3
0.600,relu,256,2,sigmoid,128,4,relu,128,3
\end{filecontents*}
\pgfplotstabletypeset[
    col sep=comma,
    skip first n=3,
    display columns/0/.style={column name = $\sigma$, /pgf/number format/.cd,fixed,precision=1},
    display columns/1/.style={column name = activ. fun., column type=|c, string type},
    display columns/2/.style={column name = h. size},
    display columns/3/.style={column name = $n$},
    display columns/4/.style={column name = activ. fun., column type=|c, string type},
    display columns/5/.style={column name = h. size},
    display columns/6/.style={column name = $n$},
    display columns/7/.style={column name = activ. fun., column type=|c, string type},
    display columns/8/.style={column name = h. size},
    display columns/9/.style={column name = $n$},
    every head row/.style={
    before row={
        \toprule
        \multicolumn{1}{c}{}& \multicolumn{3}{c}{N=500} & \multicolumn{3}{c}{N=2\thinspace000} & \multicolumn{3}{c}{N=8\thinspace000}\\
        },
      after row=\midrule,
      },
      every last row/.style={
      after row=\bottomrule},
    ]{tables/hyperparam_mlp_chen.csv}
    }\\
    \subfloat[][\textbf{LSTM}: The hyperparameters are the dropout rate (drop.),  the number of hidden layers (h. size) and the number of stacked layers (n. layers).]{
    \begin{filecontents*}{tables/hyperparam_lstm_chen.csv}
,i,i,i,i,i,i,i,i,i
n_batches,5,5,5,20,20,20,80,80,80
hyperparam,dropout,hidden size,n layers,dropout,hidden size,n layers,dropout,hidden size,n layers
noise_levels,,,,,,,,,
0.000,0.0,128,2,0.0,32,1,0.0,32,2
0.300,0.3,128,2,0.0,64,3,0.0,64,2
0.600,0.0,128,3,0.0,128,3,0.3,64,2
\end{filecontents*}
\pgfplotstabletypeset[
    col sep=comma,
    skip first n=3,
    display columns/0/.style={column name = $\sigma$, /pgf/number format/.cd,fixed,precision=1},
    display columns/1/.style={column name = drop., precision=1, column type=|c},
    display columns/2/.style={column name =  h. size},
    display columns/3/.style={column name = n. layers},
    display columns/4/.style={column name = drop., precision=1, column type=|c},
    display columns/5/.style={column name =  h. size},
    display columns/6/.style={column name = n. layers},
    display columns/7/.style={column name = drop., precision=1, column type=|c},
    display columns/8/.style={column name =  h. size},
    display columns/9/.style={column name = n. layers},
    every head row/.style={
    before row={
        \toprule
        \multicolumn{1}{c}{}& \multicolumn{3}{c}{N=500} & \multicolumn{3}{c}{N=2\thinspace000} & \multicolumn{3}{c}{N=8\thinspace000}\\
        },
      after row=\midrule,
      },
      every last row/.style={
      after row=\bottomrule},
    ]{tables/hyperparam_lstm_chen.csv}
    }
\end{table*}

\subsubsection{Silverbox}
Some hyperparameters were just experienced with manually to find good values and did not effect the results in any major fashion. For LSTM and MLP, dropout was disabled this way.

For the TCN, the kernel size was set to $2$ after initial experimentation. The number of layers (range $\{2,3\}$), the number of hidden units (range $\{4,8,16,32\}$) and dropout (range $\{0, 0.05, 0.1, 0.2\}$) were optimized using grid search. Similarly to the other experiments dropout yielded no gain and the best network had 2 layers with 8 units per layer.  Total time consumption for this optimization and hyper parameter search was 33 hours.

The MLP was implemented as a single hidden layer neural network with ReLU as activation function. The model order (range $\{1,2,4,8,16,32,64\}$) and the number of hidden units (range $\{ 4,8,16,32,64,128,256\}$) were optimized using grid search and the best hyper parameters were $4$ and $32$ respectively. Total time consumption for this optimization and hyperparameter search was 7 hours.  

In the LSTM case, the hyperparameters for batch size (range $\{1,2,4,8,16,32\}$) and the number hidden units (range $\{4, 16, 36, 64 \}$)  were optimized using grid search and the best hyper parameters were $8$ and $36$ respectively. Total time consumption for this optimization and hyperparameter search was 40 hours.

\subsubsection{F-16 ground vibration test}
Again, we used grid search for finding the hyperparameters. In each possible training configuration, we have trained the TCN for all possible combinations of: number of hidden layers in \{16, 32, 64, 128\}; dropout rate in \{0.0, 0.3, 0.5, 0.8\}; number of residual blocks in \{1, 2, 4, 8\}; for the kernel size  in \{2, 4, 8, 16\}; for the presence or absence of dilations; and, for the use of \textit{batch norm}, \textit{weight norm} or nothing after each convolutional layer. The best result in this case, is to use batch norm, no dilation, kernel size equals to $8$, dropout rate equals to $0.3$, and $8$ layers. Training the network with this configuration took approximately 40 minutes.

We have trained the MLP for all combination of: number of hidden layers in \{16, 32, 64, 128, 256\}; model order $n$ in \{2, 4, 8, 16, 32, 64, 128\}; activation function in \{ReLU, sigmoid\}. Use sigmoid, model order equals to $64$ and $256$ hidden units yields the best results. Training the network with this configuration took 4 minutes.  Training the network with this configuration took approximately 5 minutes.

Finally, we trained the LSTM for all combinations of: number of hidden layers in \{16, 32, 64, 128\}; number of stacked LSTM layers in \{1, 2\}; dropout rate in \{0.0, 0.3, 0.5, 0.8\}. The best configuration is $2$ stacked LSTM layers, dropout rate equals to $0$  and hidden size equals to 128. Training the network with this configuration took approximately 50 minutes.
\end{document}